\documentstyle[twoside,fleqn,espcrc2]{article}

\newcommand{\AmS}{{\protect\the\textfont2
  A\kern-.1667em\lower.5ex\hbox{M}\kern-.125emS}}

\input epsf
\title{Finite $ma$ corrections for sea quark matrix elements}

\author{J.-F. Laga\"e and K.-F. Liu\\
\vspace{0.25cm}
Dept. of Physics and Astronomy \\ 
Univ. of Kentucky, Lexington, KY 40506, USA}
       
\begin{document}

\begin{abstract}
We discuss the finite $ma$ corrections associated with the computation of
sea quark matrix elements. We find them to differ from the standard 
normalization used for valence quarks and to depend strongly on the Lorentz
structure of the current under consideration. Phenomenological implications 
of these results are briefly discussed in two examples. We also
mention how the magnitude of the correction factors can be reduced by using
a 2-link improved action.
\end{abstract}

% typeset front matter (including abstract)
\maketitle

Matrix elements of bilinear quark operators, measured with the Wilson fermionic action
are usually remormalized by a factor:
  \begin{equation}
  Z={2\tilde\kappa} (1+ma) [1+O(\alpha_V)]
  \end{equation}
where, in the context of mean-field improved perturbation theory \cite{Lepa93},
  \begin{equation}
  \tilde\kappa=\kappa U_0 \simeq {\kappa \over 8\kappa_c} \quad\quad
  \mbox{and} \quad\quad ma= {1 \over 2\tilde\kappa} - 4
  \end{equation} 
Here, we would like to show that this formula does not apply when the current
is part of a ``disconnected'' quark loop (as is the case for example in
computations of sea quark matrix elements or in determinations of the mass
of flavor singlet mesons). Instead, we will find that in these situations,
the factor $(1+ma)$ in (1) has to be replaced by another $ma$ dependent factor,
which depends on the Lorentz structure of the current under
consideration.

The limitations of the $(1+ma)$ factor are easily understood by noting the 
assumptions going into its derivation \cite{Lepa92}.
 It is obtained by trying to match
the lattice and continuum propagators of a {\it zero-momentum} quark
{\it on mass shell}. One can reasonably argue that this is the relevant 
situation
for a {\it valence} quark inside of a hadron at rest (although there is
room for improvement at small and intermediate values of $ma$). However, 
this is clearly insufficient to deal with the case of closed quark loops
where one has an integration over 4-momentum. This is even more so given
that, on the lattice, the integral will be sensitive to the contributions 
coming from the doublers at the edges of the Brillouin zone.
As is well known, Wilson fermions describe 16 species with masses
  $$\begin{array}{ccccc}
  ma & ma+2r & ma+4r & ma+6r &
  ma+8r \\
  \{1\} & \{4\} & \{6\} & \{4\} & \{1\}
  \end{array}$$
where the degenaracies have been indicated between braces.
When $ma \rightarrow 0$, the doublers decouple (apart from giving rise to
the anomaly) and continuum results are recovered \cite{Smit81}.
 However, in current lattice
simulations, the condition $ma \ll r$ is not always satisfied, so that the 
degeneracy between fundamental fermions and doublers is only partially lifted.
This induces lattice artifacts which are reminiscent of the naive fermion
case: closed quark loops involving a scalar current are overestimated, whereas
those associated with a pseudoscalar or an axial current are underestimated. 
As we will see below, this requires large corrections even for values of
$ma$ as low as $0.1$ (at $r=1$) which on current lattices corresponds roughly
to the mass of the strange quark. Disposing of a method for estimating the
magnitude of these renormalizations is therefore of considerable 
phenomenological importance.

The case of the pseudoscalar current was considered several years ago in an
interesting series of papers by Smit and Vink \cite{Smit87}.
 Their starting point is an
exact relation (derived from the anomalous Ward identity) between the
pseudoscalar charge ($Q_5 = \int d^4x j_5(x)$) and the topological charge
of the background gauge field ($Q_t$):
  \begin{equation}
   mQ_5=Q_t
  \end{equation} 
By imposing that this continuum relation should be satisfied for Wilson 
fermions on the lattice, they were able to deduce the appropriate $ma$
dependent renormalization factor for the lattice pseudoscalar current.
Here we would like to obtain the corresponding normalization for other 
currents, in cases where an exact relation of the type (3) is not available.
For this we will make the assumption that the $ma$ dependence of the 
renormalization factors is independent of what background field is considered,
so that it can be computed in perturbation theory for a weak external field.
The relevant diagrams at lowest order are triangle diagrams and by comparing 
their lattice and continuum values, we can deduce the appropriate 
correction factors (see \cite{Lagae94} for details).
 For example, if we consider the
case of the scalar current, we would find in the continuum the perturbative
relation:
  \begin{equation}
   m<\bar{\psi}\psi> = C( {p^2 \over m^2} ) F^2_{\mu\nu}(p)
  \end{equation}
giving the zero momentum matrix element of the scalar current in a 
background gauge field of momentum p, where $C$ is a known dimensionless
function. Doing a similar computation on the lattice, we would find:
  \begin{equation}
  m<\bar{\psi}\psi> = L(pa,ma) F^2_{\mu\nu}(p)
  \end{equation}
where the lattice dimensionless function L can now depend separately on
$pa$ and $ma$ (but will give back $C$ when both tend to 0). For simplicity,
we will consider here only the case of smooth background fields
(i.e. $p \rightarrow 0$). Then, we deduce from (4) and (5) that the appropriate
$ma$ dependent correction factor for the scalar current is:
  \begin{equation}
  \kappa_S(ma) = C(0)/L(0,ma)
  \end{equation} 
Similar computations can be done for other currents (in particular, for
the pseudoscalar current, one recovers the result of Smit and Vink
mentioned above). The results are given in fig. 1 for the 
scalar ($\kappa_S$), axial ($\kappa_A$) and pseudoscalar ($\kappa_P$) 
currents (to increase the lisibility, the inverse of the
renormalization factors have been plotted). 
\begin{figure}[htb]
\vspace{-1.2cm}
\epsfxsize=7.5cm
\epsfbox{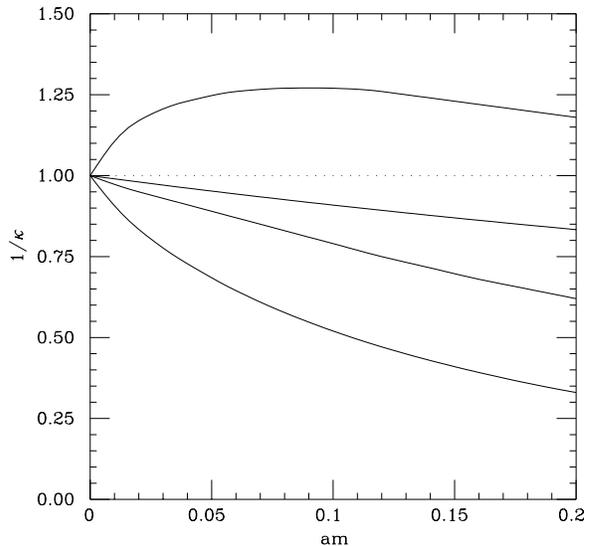}
\vspace{-2cm}
%\framebox[55mm]{\rule[-21mm]{0mm}{43mm}}
\caption{ {\it Inverse} normalization factors for various sea quark currents 
versus quark mass. The valence factor has been included for comparison
[$\kappa^{-1}_V = 1/(1+ma)$]. From top to bottom: scalar, valence, axial 
and pseudoscalar}
\label{fig:largeenough}
\end{figure}
We would like to reiterate that those factors
are to replace the ($1+ma$) part of the renormalization factor whenever
``disconnected'' quark loops are involved. They merely correct for the
different ``response function'' of the lattice and continuum fermionic
currents to the same background gauge field. They differ from current to
current because of the different incidence of the doubling problem in 
different channels. The rest of the renormalization factor
$[1+O(\alpha_V)]$ is as usual.
It appears that these correction factors are rather large and will therefore
strongly influence the interpretation of simulation results. To illustrate 
this, we now discuss their application to two interesting problems.
The first is the computation of the mass of the $\eta^\prime$. In a quenched
simulation, it can be estimated from the ratio of a 2-loop to 1-loop 
amplitude \cite{Kura94}.
 As a consequence of the above discussion, we expect that they
will have different renormalization factors: The 1-loop part behaves as a 
``valence'' quark (and will therefore pick-up a ($1+ma$) correction for
each current) whereas the 2-loop diagram corresponds to a ``sea'' quark
situation (in which case, the pseudoscalar factor of fig. 1 is relevant).
\begin{figure}[htb]
\vspace{-1.2cm}
\epsfxsize=7.5cm
\epsfbox{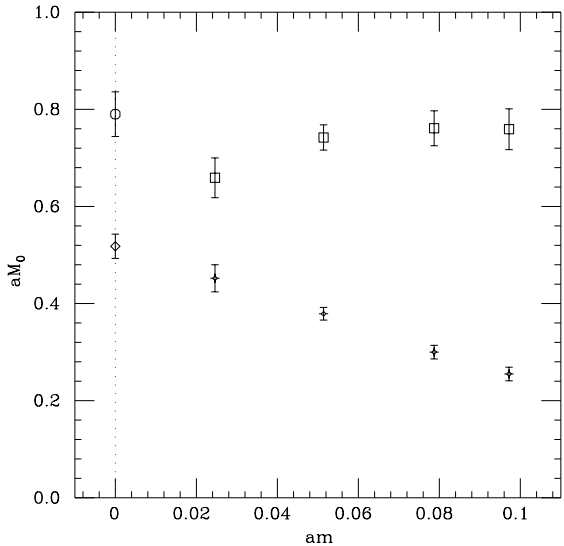}
\vspace{-2cm}
\caption{$\eta_0$ mass vs quark mass after renormalization (squares) and 
before (plusses). Also included are the chiral extrapolation of the raw
data (diamond) and the estimate from the 
Witten-Veneziano formula (circle), as obtained in $[6]$.}
\label{fig:toosmall}
\end{figure}
In figure 2, we compare the mass of the $\eta^\prime$ after correction with
the raw lattice data \cite{Kura94}.
 The trends are very different: after correction,
the mass of the $\eta^\prime$ is essentially independent of the mass of the
quark up to the strange quark mass with a chiral limit that is in much better
agreement with the estimate obtained from the Witten-Veneziano formula 
(We assume that the drop in the renormalized data
at low quark masses can be attributed to finite size effects or to the zero-mode
shift effect \cite{Smit87}). The agreement between the renormalized data and
the Witten-Veneziano estimate doesn't come as a surprise,  
since the correction factor that we are advocating
here is the same as the one used by Smit and Vink \cite{Smit87} to compute
 the topological
charge by the fermionic method. As a second example, we mention the 
contribution of the strange quark to the mass of the nucleon. If we replace
the ($1+ma$) correction factor used in \cite{Fuku94}, with the sea quark scalar correction
factor from fig. 1, we find that the strange quark contribution to the mass
of the nucleon comes down, from $30\%$ to $21\%$.

Since the correction factors given above are generally large, there is much 
interest in trying to reduce their magnitude by using modified actions. 
We found in \cite{Lagae94} that for the axial and pseudoscalar current 
considerable improvement can be achieved by using the 2-link action of 
Hamber and Wu \cite{Wu84}. For the scalar current, on the other hand, 
this kind of action doesn't seem to help much and it is probably more 
interesting 
to consider other alternatives, such as staggered fermions for example or 
working with naive fermions and dividing the final result by 16 \cite{Lagae94}.
 Finally, we would like to recall that the renormalization factors 
presented here have been 
obtained by assuming a lattice simulation dominated by low 
momentum gauge fields. In more general situations it might be necessary to 
consider the ``response functions'' at finite momentum \cite{Lagae94}. 
This however is left for later studies.

\end{document}